\documentclass[aps,twocolumn,showpacs]{revtex4}
\usepackage{epsfig}
\newcommand{\be}{\begin{equation}}
\newcommand{\ee}{\end{equation}}
\begin{document}

\title{Comment on ``Central limit behavior in deterministic dynamical systems"}

\author{Peter Grassberger}
\affiliation{
Complexity Science Group, Department of Physics and Astronomy, University of Calgary,
Calgary, Canada}

\date{\today}

\begin{abstract}
We check claims for a generalized central limit theorem holding at the 
Feigenbaum (infinite bifurcation) point of the logistic map, made recently by 
U. Tirnakli, C. Beck, and C. Tsallis (Phys. Rev. {\bf 75}, 040106(R) (2007)). 
We show that there is no obvious way that these claims can be made consistent 
with high statistics simulations. We also refute more recent claims by the 
same authors that extend the claims made in the above reference.
\end{abstract}

\maketitle

In \cite{tirnakli}, the authors claim that it is ``less known in the physics community"
that the central limit (CL) theorem holds for deterministic but mixing 
dynamical systems. Our first comment is that this connection is perfectly well known 
and was never doubted. An early reference is \cite{GN}.

But my main criticism concerns the treatment of the (non-mixing) logistic map at the 
Feigenbaum (infinite bifurcation \cite{Feigen}) point in \cite{tirnakli}. Let us consider 
trajectories $x_{i+1} = f_a(x_i)$ of length $N$, with $f_a(x) = a-x^2$. 
The Feigenbaum point is at $a=a_c\equiv 1.40115518909\ldots$. Following \cite{tirnakli},
we study sums $Y = \sum_{i=N_0+1}^{N_0+N} x_i$ and their distributions for random $x_0$
at $a=a_c$ and for large $N$. Here, $N_0$ is the length of a possible discarded 
transient. At first we shall consider the case $N_0=0$, i.e. no transient is 
discarded. We understand that this is also the case studied in \cite{tirnakli}. At 
least the authors do not mention any transient, although there are reasons (discussed later)
to suspect that that they might have used $N_0>0$.

Denoting by $\langle Y\rangle$ the average over $x_0$, the claim of 
\cite{tirnakli} is that the centered and suitably rescaled sums
\be
   y = N^{-\gamma} (Y - \langle Y\rangle)\;,
\ee
are distributed according to a ``$q-$Gaussian"
\be
   p(y) \propto {1\over [c+y^2]^b} 
\ee
for $\gamma=1.5$, with $b\approx 4/3$ and $c\approx 0.1$ \cite{footnote}. Moreover, it 
is claimed that 
the same distribution, with identical $\gamma, b$, and $c$, is found for the modified 
logistic map $f_{a,z}(x) = a-x^z$ with $z=1.75$ and $z=3$. If true, this universality 
would be remarkable.
	
\begin{figure}
\psfig{file=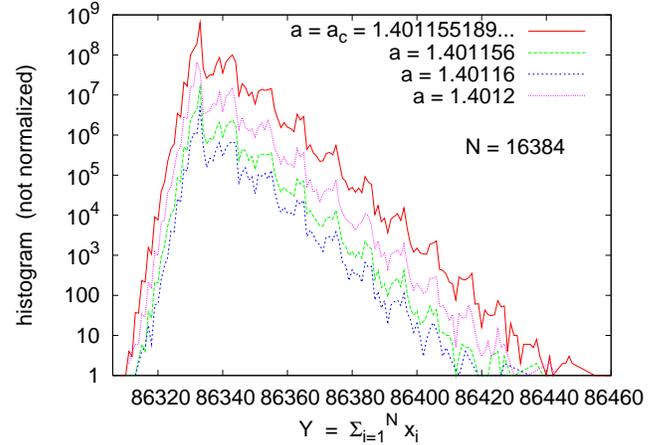,width=6.0cm,angle=270}
\caption{Histograms of $Y$ for $N=16384$ and for four values of $a$ at or slightly 
above criticality. }
\end{figure}


\begin{figure}
\psfig{file=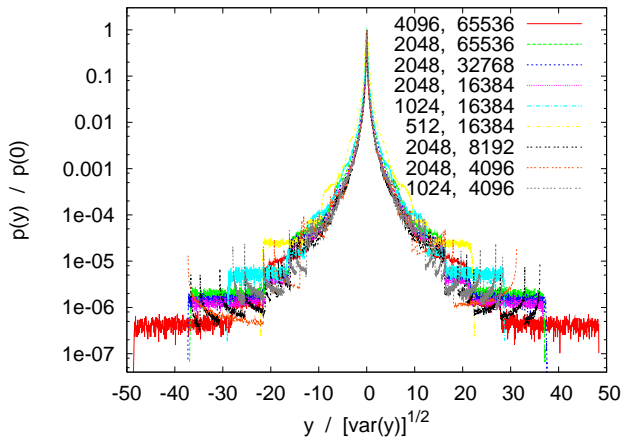,width=6.0cm,angle=270}
\caption{Distributions of $y/ \sqrt{var(y)}$, normalized to $p(0)=1$, for various 
values of $N$ and $n$, where $a$ is set to the $n\to n-1$ band merging point. In 
all cases, $N_0 \geq 16384$. Similar results were obtained also for other values 
of $a$. Notice that the statistics in any curve of this figure is at least 10 times higher 
than in any of the curves in \cite{tirnakli,tirnakli2}}.
\end{figure}

\begin{figure}
\psfig{file=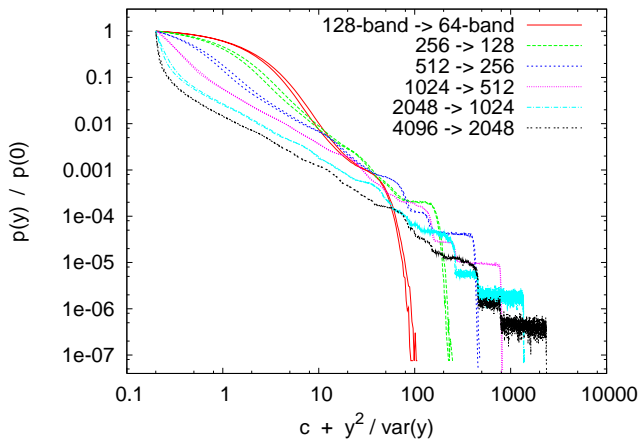,width=6.0cm,angle=270}
\caption{Distributions of $y / \sqrt{var(y)}$ for $N=65536$, normalized 
to $p(0)=1$, plotted against $y^2+const$ 
on a log-log plot. As in Fig.~2, long transients ($N_0 = 65536$ in most cases) have 
been discarded, and the control parameter $a$ of the logistic 
map is chosen as the $n\to n/2$ band merging point.
According to Eq.~(2), one would expect straight lines with slopes
$-b$. Apart from the rather unsystematic deviations at large $y$ which could be 
effects which vanish in the limit $N\gg n\to\infty$, one sees a systematic downward 
curvature for intermediate $y$ and strong systematic upward (downward) curvatures 
for $n>\sqrt{N}$ ($n<\sqrt{N}$) at very small $y$. Notice that one has two curves for 
each $n$, one for $y>0$ and one for $y<0$.}
\end{figure}

Unfortunately, none of the above claims seem to be correct. It is straightforward to 
do the necessary simulations to estimate $p(y)$. In all simulations, $x_0$ was uniformly
distributed in $[0,a]$. Results, for several values of $a$ 
at and slightly above $a_c$, are shown in Fig.1. Indeed, in this figure are shown 
histograms of the non-rescaled and non-shifted sums $Y$, for $N=16384$. Similar 
results were obtained for other values of $N$. They have markedly different 
behavior left and right of the central peak $Y_c\approx 86333$. For $Y<Y_c$ we observe 
a very steep rise, $P(Y) \sim e^{0.86Y}$, while the decrease for $Y>Y_c$ is much more gentle,
$P(Y) \sim e^{-0.17Y}$. Superposed on both exponentials are periodicities which 
obviously result from the hierarchical structure of the Feigenbaum attractor.

In obtaining Fig.1 we have not discarded any transient ($N_0=0$). 
Notice that the entire dynamics is transient 
at the Feigenbaum point \cite{Feigen,GS}. Thus discarding a finite transient 
would introduce a new time scale and ruin any hope for scaling. 

This would be different, if values $a>a_c$ were considered as in \cite{tirnakli2} 
and, presumably, also in \cite{tirnakli}: The results of Ref.\cite{tirnakli2} are 
most easily understood as artifacts generated by using the 9-digit approximation 
1.40115519 for $a_c$. 
Let us consider a value of $a$ where the attractor consists of $n=2^k$ ``bands"
\cite{Feigen,GS}. Orbits on it jump periodically between the bands, but are chaotic
within each band.
Then it makes sense to discard transients of length $\gg n$. On each band, the 
$n-$fold iterated map $f^{(n)}_a$ is mixing. Thus $Y$ is a sum over $n$ series 
of random variables, 
each of which shows normal CL behavior for $N\to\infty$. Therefore, 
$Y$ also shows normal CL behavior in the limit $N\to\infty,\;n=const$. 

Deviations from normal CL behavior can be expected only when taking a joint
limit $a\to a_c$ (i.e., $n\to\infty$) and $N\to\infty$. 
Indeed, in \cite{tirnakli2} it was proposed that $N\sim n^2$. 
Notice, however, that in this case we are no longer dealing with the problem posed 
in the central limit theorem, i.e. the asymptotics of partial sums in an infinite 
sequence of random variables. Thus, strictly speaking, we are no longer dealing 
with (normal or abnormal) central limit behavior at all.

In the following we shall, for definiteness, only deal with band-merging points, 
where $n=2^k$ bands merge into $n/2$ bands as $a$ is increased. But similar 
behavior is found also for other values of $a$.
Indeed, when looking at distributions of $y$ for large $n$, $N>>n$, and $N_0>>n$, 
one finds heavy-tailed distributions, see Fig.~2. But as closer inspection shows,
they are in general not described by Eq.~(2) (see Fig.~3). Apart from the steps 
and discontinuities at large $y$ which might recede to infinity in the limit 
indicated above, the main deviations are:
\begin{itemize}
\item A systematic downward curvature in Fig.~3 for intermediate to large $y$;
\item Deviations from straight line behavior at very small $y$, both for 
$n<<\sqrt{N}$ and for $n>>\sqrt{N}$. If at all, the data are compatible for small 
$y$ with Eq.~(2) only for a very narrow region of $N/n^2$. 
\end{itemize}
In view of this it seems very unlikely that the rough agreement with Eq.~(2) is 
more than a numerical coincidence. The data shown in Figs.~4 and 5 of \cite{tirnakli}
are definitly not well fitted by Eq.~(2) (not even for very small $y/\sigma$). 
We might add that equally good (or bad) fits 
would be obtained with Levy stable distributions \cite{levy}, which moreover have more 
theoretical justification.

Some final remarks:
\begin{itemize} 
\item The behavior described here is seen only when $N$ is a power of 2.
Otherwise, one observes completely different behavior.
\item The fluctuations of $Y$ are, for $N_0, N \gg n \gg 1$, 
tiny. All structures shown in Figs.~2 and 3 (including the tails!) extend, before 
centering and dividing by $N^\gamma$, over a range $\Delta Y < 10^{-3}$. For $N=65536$, 
this is to be compared to $\langle Y \rangle \approx 34533$, i.e. all relative fluctuations 
are smaller than $3\times 10^{-8}$ \cite{footnote2}. 
The reason for this is that the motion on an $n-$band 
attractor with large $n$ is extremely regular, with the chaos confined to very narrow 
bands. Thus if a generalized CL theorem holds for this problem in any sense, it is 
completely unobservable in any experimental situation.
\item Since the phenomenon 
illustrated in Figs.~2, 3 seems to describe corrections to the scaling limit of the 
Feigenbaum map, it is not clear how much it depends on the original map one starts
from and on the distribution of $x_0$. The only phenomenon discussed in this comment 
which has a realistic chance to be experimentally accessible 
and is likely to be universal is the behavior shown in Fig.~1. It is dominated by chaotic
transients, and is very far from anything described in Refs.~\cite{tirnakli,tirnakli2}.
\end{itemize}


\begin{thebibliography}{99}
\bibitem{footnote} In the $q-$statistics community, Eq.~(2) is usually written in a more 
cumbersome way. We prefer the simpler and more intuitive form given here.
\bibitem{footnote2} This also raises the question of numerical precision.
All simulations in the present comment used the GCC compiler on an Intel platform, 
with floating point operations performed with ``long double" (80 bit)
data type. Tests were made with standard double precision (64 bit).
\end{thebibliography}
\end{document}